\documentclass[twocolumn,prl,superscriptaddress]{revtex4}
\usepackage{graphicx}
\usepackage{dcolumn}
\usepackage{amsmath,amsthm,amssymb}

\begin{document}
\title{Codimension-two critical behavior in vacuum gravitational collapse}

\author{Piotr Bizo\'n}
\affiliation{M. Smoluchowski Institute of Physics, Jagiellonian
University, Krak\'ow, Poland}
\author{Tadeusz Chmaj}
\affiliation{H. Niewodniczanski Institute of Nuclear
   Physics, Polish Academy of Sciences,  Krak\'ow, Poland}
   \affiliation{\small{Cracow University of Technology, Krak\'ow,
    Poland}}
\author{Bernd G. Schmidt}
    \affiliation{Max-Planck-Institut f\"ur Gravitationsphysik,
Albert-Einstein-Institut, Golm, Germany}
\date{\today}
\begin{abstract}
We consider the critical behavior at the threshold of black hole
formation for the  five dimensional vacuum Einstein equations
satisfying the cohomogeneity-two triaxial Bianchi IX ansatz.
Exploiting a discrete symmetry present in this model we predict the
existence of a codimension-two attractor. This prediction is
confirmed numerically and the codimension-two attractor is
identified as a discretely self-similar solution with two unstable
modes.

\end{abstract}

\maketitle

\emph{Introduction.}
Since the pioneering work of Choptuik  on the collapse of
self-gravitating scalar field \cite{ch}, the nature of the boundary
between dispersion and black hole formation in gravitational
collapse  has been a very active research area (see \cite{g} for a
review). One of the most intriguing aspects of these studies is the
occurrence of discretely self-similar critical solutions. Discrete
self-similarity (DSS) means invariance under rescalings of space and
time variables by a constant factor $e^{\Delta}$ where  $\Delta$ is
a number, usually called the echoing period.
   Critical solutions possessing this curious symmetry (with
different echoing periods) have been found for several
self-gravitating matter models (massless scalar field \cite{ch},
Yang-Mills field \cite{ccb}, $\sigma$-model \cite{vienna} and few
others) and recently also in the vacuum gravitational collapse in
higher dimensions \cite{bcs, bcs2}.

Our current understanding of DSS solutions is very limited in
comparison with  continuously self-similar (CSS) solutions. In the
case of spherical symmetry the CSS ansatz leads to an ODE system
which can be handled analytically and sometimes even rigorous proofs
of existence are feasible. For example, the existence of a countable
family of CSS solutions was proved in the Einstein-sigma model
\cite{bw} and the ground state of this family was identified as the
critical solution (which was known previously from numerical studies
of  critical collapse).  In contrast, the DSS ansatz leads to a
$1+1$ PDE eigenvalue problem which seems intractable analytically.
Although this eigenvalue problem can be solved numerically, as was
done by Gundlach for two models (scalar field \cite{g1} and
Yang-Mills \cite{g2}),  the numerical iterative procedure requires a
good initial seed in order to converge. Thus, Gundlach's method is
efficient in validating and refining DSS solutions which are already
known from direct numerical simulations but it is not useful in
searching for new solutions.

  In this paper we consider the critical collapse for the five dimensional vacuum
  Einstein equations
  satisfying the cohomogeneity-two triaxial Bianchi IX ansatz and provide
  heuristic arguments and numerical evidence for the
   the existence of
  a  DSS solution with two unstable modes.  This is a continuation of our studies in \cite{bcs}
  where we have shown the existence of a critical DSS
  solution with one unstable mode and the associated type II
  critical behavior in this model.
On the basis of our result it is tempting to conjecture that the
critical DSS solution is a ground state of a countable family of DSS
solutions with increasing number of unstable modes.
\vskip 0.2cm \emph{Background.} Our starting point is the
cohomogeneity-two symmetry reduction of the Einstein equations in
five dimensions which is based on the following ansatz introduced by
us in \cite{bcs}
\begin{equation}\label{metric}
    ds^2\!=\! - A e^{-2\delta} dt^2 + A^{-1} dr^2 + \frac{1}{4} r^2 \left[ e^{2B}
    \sigma_1^2\!+\!e^{2C} \sigma_2^2\! +\!e^{-2(B+C)}\sigma_3^2\right],
\end{equation}
where $A$, $\delta$, $B$, and $C$ are functions of time $t$ and
radius $r$. The angular part of (\ref{metric})  is the
$SU(2)$-invariant homogeneous metric on the squashed 3-sphere with
$\sigma_k$ being standard left-invariant one-forms on $SU(2)$
\begin{equation}\label{sigma}
    \sigma_1+i\,\sigma_2=e^{i\psi} (\cos{\theta} \;d\phi + i\,
    d\theta), \quad \sigma_3=d\psi - \sin{\theta}\; d\phi.
\end{equation}
where  $0\leq \theta \leq \pi, 0\leq \phi \leq 2\pi$, $0\leq \psi
\leq 4\pi$ are the Euler angles. The squashing modes, $B$ and $C$,
play the role of dynamical degrees of freedom. This ansatz provides
a simple $1+1$ dimensional framework for investigating the dynamics
of gravitational collapse in vacuum. In \cite{bcs} we made a
simplifying assumption that $B=C$ which means that the ansatz
(\ref{metric}) has an additional $U(1)$ symmetry and only one
dynamical degree of freedom (so called biaxial case). In this paper
we drop this assumption and consider the full triaxial case with two
dynamical degrees of freedom.

 Substituting the ansatz
(\ref{metric}) into the vacuum Einstein equations in five dimensions
and using the mass function $m(t,r)$, defined by $A=1-m/r^2$, we get
the following system \begin{widetext}
\begin{subequations}
\begin{eqnarray}
m' &=& \frac{2}{3} r^3\left[e^{2\delta} A^{-1} ({\dot B}^2 +{\dot
C}^2+\dot B
\dot C)+ A ({B'}^2+{C'}^2+B' C')\right] \nonumber \\
&&\hskip 0.9cm+\frac{2}{3} \,r
 \left(3+ e^{4B}+ e^{4C}-2e^{-2B}-2e^{-2C}-2e^{2(B+C)}+e^{-4(B+C)}\right)\,,\\
\dot m & = & \frac{2}{3} r^3 A \left(\dot C B' + \dot B C' + 2\dot B
B' + 2\dot C C'\right)\,, \\
  \delta' &=& - \frac{2}{3}\, r \left[e^{2\delta} A^{-2}({\dot B}^2 +{\dot
  C}^2 +\dot B \dot C)+
    B'^2+C'^2+B' C'\right]\,,\\
 \left(e^{\delta} A^{-1} r^3 {\dot B}\right)^{\cdot} &=&
\left(e^{-\delta} A r^3 B'\right)' \!- \frac{4}{3} e^{-\delta} r
\left(2 e^{4B}\!+\!2
e^{-2B}\!-\!e^{2(B+C)}\!-\!e^{-4(B+C)}\!-\!e^{4C}\!-\!e^{-2C}\right)\,,\\
 \left(e^{\delta} A^{-1} r^3 {\dot C}\right)^{\cdot} &=&
\left(e^{-\delta} A r^3 C'\right)'\! - \frac{4}{3} e^{-\delta} r
\left(2 e^{4C}\!+\!2
e^{-2C}\!-\!e^{2(B+C)}\!-\!e^{-4(B+C)}\!-\!e^{4B}\!-\!e^{-2B}\right)\,,
\end{eqnarray}
\end{subequations}
where primes and overdots denote derivatives with respect to $r$ and
$t$, respectively.
\end{widetext}

If $B=C=0$, then the ansatz (\ref{metric}) reduces to the standard
spherically symmetric metric for which the Birkhoff theorem applies
and the only solutions are Minkowski ($\delta=0, m=0$) and
Schwarzschild ($\delta=0, m=const>0$). We showed in the biaxial case
\cite{bcs} that these two static solutions play the role of
attractors in the evolution of  generic regular initial data  (small
and large ones,
  respectively). We have verified that this property remains true in the triaxial
  case (see also \cite{dh}).
Note that equations for small perturbations $\delta B$ and $\delta
C$ around the Minkowski and Schwarzschild solutions decouple, hence
linear stability results and the decay rates discussed in \cite{bcs}
carry over without any changes.

\vskip 0.2cm \emph{Heuristics.} Below we focus on initial data on a
borderline between dispersion and collapse.  Our preliminary studies
of the evolution of such  data suggested that the relaxation of
symmetry from biaxial to triaxial does not change the phenomenology
of critical behavior observed in \cite{bcs}, that is we have found
our old biaxial DSS solution
 (hereafter referred to as the $DSS_1$)  as the critical solution. In other words the
biaxial symmetry seems to be recovered dynamically not only for
generic initial data but also for the critical ones (see Fig.~1).
The mechanism of this
 nonlinear synchronization phenomenon is an open problem
 which we hope to pursue elsewhere. Here we take it as a fact and
 use it as the starting point in further discussion.
\begin{figure}[h]
\centering
\includegraphics[height=0.4\textwidth,width=0.5\textwidth]{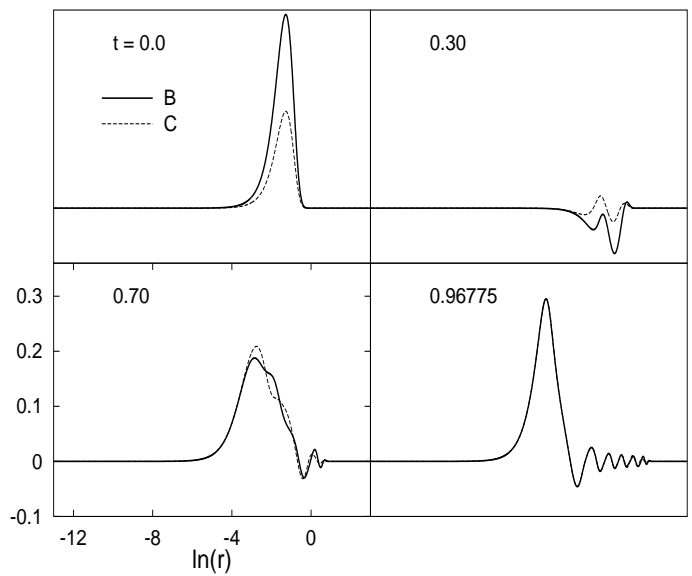}
\caption{\small{Snapshots of the evolution of near-critical initial
data of the form (\ref{ic}) for $a=1/2$. In the last frame the
squashing modes become synchronized and the solution coincides with
the $DSS_1$. }\vspace{-0.3cm}}\label{fig1}
\end{figure}

  The second key element of our
 argument is the fact that the system (3) possesses a discrete $Z_3$ symmetry
 which
corresponds to the freedom of permutations of coefficients of
one-forms $\sigma_k$ in the angular part of the metric
(\ref{metric}). These permutations are generated by the following
transpositions
\begin{eqnarray}\label{trans}
   &&T_{12}:(B,C)\rightarrow (C,B),\quad T_{23}:(B,C)\rightarrow (B,
   -B-C),\nonumber\\
   &&T_{13}:(B,C)\rightarrow (-B-C, C)\,,
\end{eqnarray}
where the transposition $T_{ij}$  swaps the coefficients of
$\sigma_i^2$ and $\sigma_j^2$ in (\ref{metric}).
Biaxial configurations correspond to the fixed points of these
transpositions: $(B,B)$, $(B,-B/2)$, and $(B,-2B)$.

Thus, each biaxial solution, in particular the $DSS_1$, exists in
three different but geometrically equivalent copies. Let
$\mathcal{M}_{crit}$ denote the codimension-one critical surface in
the phase space which separates dispersion from collapse  and let
$\mathcal{M}_i$ ($i=1,2,3$) denote the basins of attraction of three
symmetry-related copies of the $DSS_1$ solution. Since
$\mathcal{M}_i$ lie in $\mathcal{M}_{crit}$ and $\mathcal{M}_{crit}$
is connected, there should exist codimension-two boundaries that
separate different $\mathcal{M}_i$ from each other. It is natural to
expect that these boundaries are given by the stable manifolds of
three symmetry-related copies of a solution with two unstable modes.
In the next section we confirm this picture numerically and
demonstrate that the codimension-two attractor is discretely
self-similar.
\vskip 0.2cm \emph{Numerics.} In order to find a codimension-two
attractor predicted above we consider the  two-parameter family of
time-symmetric initial data of the form
\begin{equation}\label{ic}
    B(0,r)=p\, f(r), \quad C(0,r)=a\,
    B(0,r),
\end{equation}
where $f(r)$ is a smooth localized  function  satisfying regularity
conditions at $r=0$. The results presented below were produced for
the generalized gaussian $f(r)=100 \,r^2 \exp(-20(r-0.1)^2)$. We
pick some value of the parameter $a$ and then, using bisection, we
fine-tune the parameter $p$ to the critical value $p^*(a)$. We are
interested in how the phenomenology of critical behavior depends on
$a$. There are three distinguished values of $a$ which correspond to
three biaxial configurations: $a=1$, $a=-1/2$, or $a=-2$.  Since
biaxiality is preserved during evolution,  for these special initial
data we obviously find our old $DSS_1$ as the critical solution
(note that the corresponding critical values of $p$ are related by
the symmetry (\ref{trans}): $p^*(-1/2)=-2p^*(1)$ and
$p^*(-2)=p^*(1)$ which provides a useful test of the accuracy of
numerics). As we have already mentioned, numerical studies of
critical collapse indicate that for other (generic) values of the
parameter $a$ the $DSS_1$ solution, modulo the $Z_3$ symmetry, also
acts as an attractor. Let us denote by $X^{(1)}_1$, $X^{(1)}_2$, and
$X^{(1)}_3$ the three copies of the $DSS_1$ solution, corresponding
to $a=1$, $a=-1/2$ and $a=-2$, respectively. Below we  focus our
attention on $X^{(1)}_1$ and $X^{(1)}_2$.  By continuity, the
solution $X^{(1)}_1$ is the attractor for the values of $a$ close to
$1$ and the solution $X^{(1)}_2$ is the attractor for the values of
$a$ close to $-1/2$. Thus, in the interval $-1/2<a<1$ there must
exist at least one critical value $a^*$ such that the critical
initial data (\ref{ic}) corresponding to $a^*\pm \epsilon$ (for a
sufficiently small $\epsilon$) evolve to the different copies
$X^{(1)}_1$ and $X^{(1)}_2$. As usual, this critical value $a^*$ can
be determined by bisection~\footnote{Actually, it turns out that
there are many (perhaps infinitely many) critical values of $a$,
intermingled in a complicated way. This suggests that the structure
of codimension-two boundaries between different basins of
attractions $\mathcal{M}_i$ is very complex, and quite likely
fractal.}. We shall refer to initial data with parameters
$(a^*,p^*(a^*))$ as double critical. In Fig.~2 we show the evolution
of near  double critical initial data. The key new feature which is
apparent in Fig.~2 (in contrast to Fig.~1) is the occurrence of
another intermediate attractor around which the solution hangs for a
while before approaching the $DSS_1$ attractor. We find that the new
attractor is discretely self-similar with the echoing period
$\Delta_2\approx 0.395$ (see Figs.~3 and ~4).  We  call this
solution the $DSS_2$ and denote by $X^{(2)}$.
\begin{figure}[h]
\centering
\includegraphics[height=0.4\textwidth,width=0.5\textwidth]{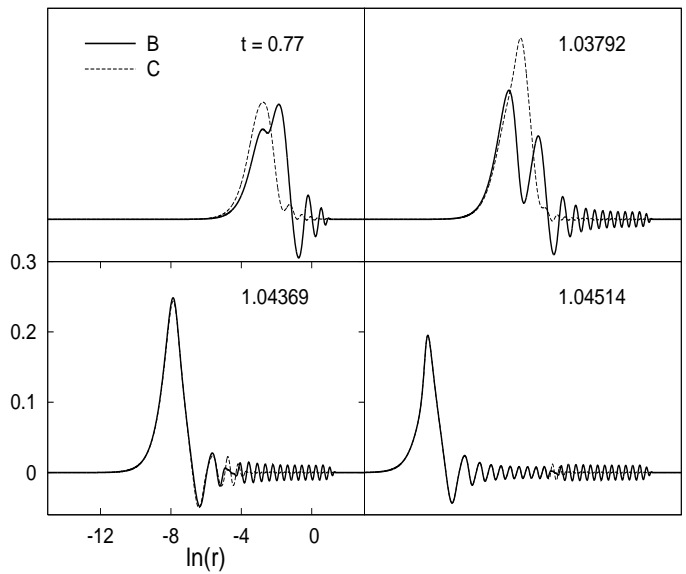}
\caption{\small{Snapshots of the evolution of near double critical
initial data ($a=0.1411...,p=0.0952...$). The new (codimension-two)
intermediate attractor ($DSS_2$) is seen in the second frame. The
departure from this attractor (proceeding from the origin) and the
approach to the old (codimension-one) intermediate attractor
($X^{(1)}_1$ copy of the $DSS_1$)  in seen in the third frame. In
the last frame the $DSS_1$ regime is well developed near the origin
while the $DSS_2$ regime is still visible for large $r$. To produce
this figure we had to fine-tune $p$ to the critical value with a
relative accuracy of $10^{-26}$ (using quadruple precision
arithmetics). } \vspace{-0.5cm}}\label{fig2}
\end{figure}
\begin{figure}[h]
\centering
\includegraphics[width=0.5\textwidth]{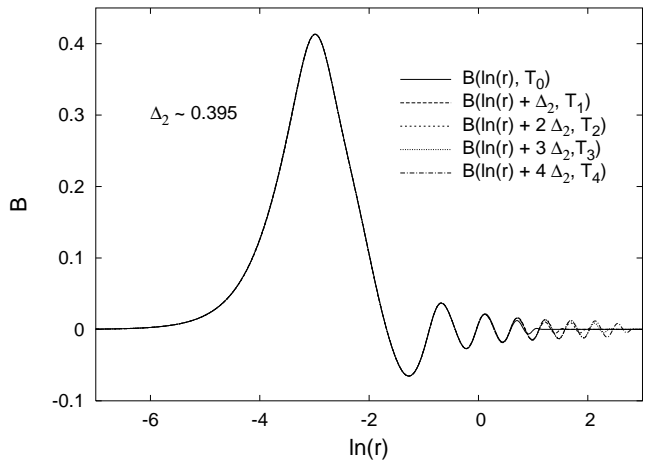}
\caption{\small{Evidence for discrete self-similarity of the
codimension-two attractor. We plot the mode $B$ at some central
proper time $T_0$ during the intermediate regime shown in the second
frame in Fig.~2 and superimpose the next four echoes which
subsequently develop.}\vspace{-0.4cm} }\label{fig3}
\end{figure}
\begin{figure}[h]
\centering
\includegraphics[width=0.5\textwidth]{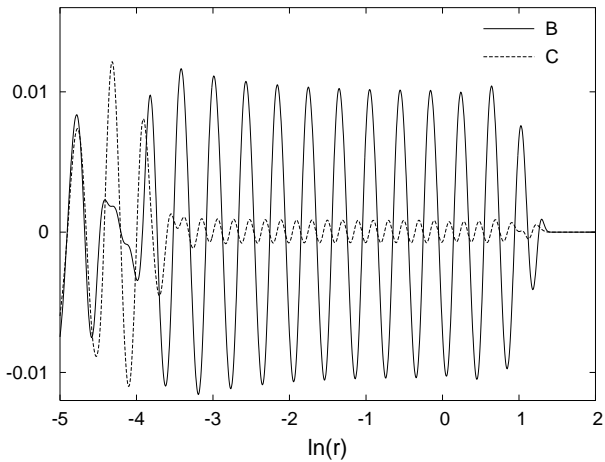}
\caption{\small{The close-up of the $DSS_2$ solution. It is evident
that this solution is  not biaxial. Note that the mode $C$
oscillates with twice the frequency of the mode $B$.}
\vspace{-0.4cm}}\label{fig4}
\end{figure}

The behavior seen if Fig.~2 has a natural explanation within the
heuristic picture sketched above. According to this picture the
$DSS_2$  has two unstable modes, one tangential  and one transversal
to the critical surface. During the $DSS_2$ intermediate regime the
departure from $X^{(2)}$ is well approximated by a linear
combination of two unstable modes
\begin{equation}
\delta X^{(2)} \!\approx c_1(a,p) e^{\lambda^{(2)}_1 \tau}
f_1(\rho,\tau) \!+\! c_2(a,p) e^{\lambda^{(2)}_2 \tau}
f_2(\rho,\tau)\,,
\end{equation}
where $\rho=r/(T-t)$ and $\tau=-\ln(T-t)$ are the similarity
variables  ($T$ is the blowup time of the $DSS_2$), $f_i(\rho,\tau)$
are the eigenfunctions (periodic in $\tau$ with the period
$\Delta_2$) and $\lambda^{(2)}_i>0$ are the respective eigenvalues.
The coefficients $c_i$ are the only vestige of initial data.  Given
$a$ close to $a^*$, for the perfect fine-tuning of $p$ to its
critical value, the transversal unstable mode (let it be $f_1$) is
completely suppressed, that is $c_1(a,p^*(a))=0$. In practice, we
can fine-tune $p$ with very high precision so that $c_1(a,p(a))$ is
almost zero. For such data $c_1 \ll c_2$, hence after several cycles
the solution leaves the $DSS_2$ along the tangential unstable
direction and then, after a short transient period, approaches one
of the copies $X^{(1)}_1$ or $X^{(1)}_2$ (depending on the sign of
$c_2$) of the $DSS_1$. Afterwards the solution  stays near $DSS_1$
for some time and eventually moves away  from the critical surface
$\mathcal{M}_{crit}$  towards collapse or dispersion.

The existence of the $DSS_2$ solution  has an interesting
consequence for the black-hole mass scaling law, $M_{BH} \sim
(p-p^*)^{\gamma}$, associated with the type II critical collapse.
Consider a near critical value $a$ and monitor the black hole mass
over a wide range of supercritical values $p>p^*(a)$. As we have
discussed above, for $p$ very close to the threshold the solution
moves away from the critical surface along the unstable mode of the
$DSS_1$. However, as $p-p^*$ increases, we observe a competition
between the transversal and the tangential modes of the $DSS_2$. Far
from the threshold the transversal mode becomes dominant ($c_1\gg
c_2$) which means that the solution moves away from the critical
surface along the transversal unstable direction of the $DSS_2$ and
never approaches the $DSS_1$. According to the well-known argument
from dimensional analysis \cite{g} the scaling exponent $\gamma$ is
related to the eigenvalue $\lambda$ of the unstable mode along which
the solution is ejected towards collapse: $\gamma=2/\lambda$ (the
factor of $2$ comes from the fact  that in five dimensions mass has
dimension $length^2$). Thus, for near double critical initial  data
the mass scaling law is expected to exhibit the crossover from the
exponent $\gamma_1=2/\lambda^{(1)}$ (where $\lambda^{(1)}$ is the
eigenvalue of the single unstable mode of the $DSS_1$) for small
$p-p^*$ to $\gamma_2=2/\lambda^{(2)}_1$ for relatively large $p-p^*$
(see Fig.~5 for the numerical confirmation of this prediction).
Using this relationship and the numerically determined $\gamma_2$ we
estimate that $\lambda^{(2)}_1\approx 6.67$.  The second eigenvalue
$\lambda^{(2)}_2$ associated with the tangential unstable mode of
the $DSS_2$ can be obtained from the relationship $\lambda^{(2)}_2
\Delta_2 \delta n = -\delta\pi$, where $\delta n$ and $\delta \pi$
are the changes in the echo number $n$ and $\pi=\ln|p-p^*|$. Fitting
this formula to numerical data we obtain $\lambda^{(2)}_2 \approx
6.03$.
\begin{figure}[h]
\centering
\includegraphics[width=0.5\textwidth]{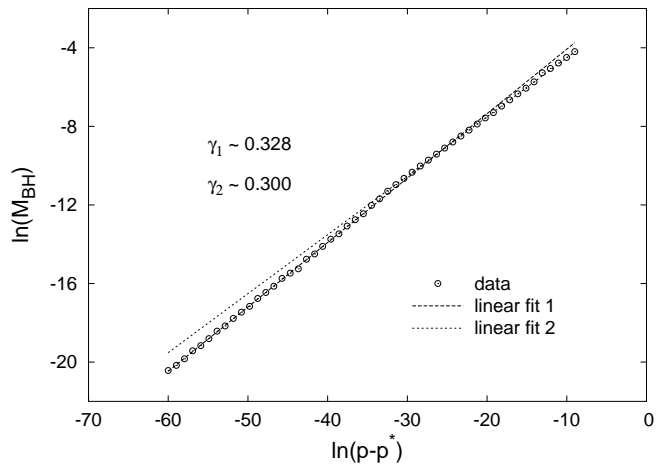}
\caption{\small{The black-hole mass scaling law for near double
critical initial data. The linear fit yields different slopes for
small and large distances from the threshold.}
\vspace{-0.3cm}}\label{fig5}
\end{figure}

We remark that the  heuristic argument for the existence of a
codimension-two attractor presented in this paper can be applied
repeatedly to argue for the existence of higher codimension
attractors~\footnote{Cf. K. Corlette and R. M. Wald, Comm. Math.
Phys. \textbf{215}, 591 (2001),
 where a similar in spirit argument is used
to prove the existence of infinitely many harmonic maps between
n-spheres.}, however the numerical search for these solutions via
multi-parameter fine-tuning would be extremely difficult.

\vskip 0.1cm \noindent {\emph{Acknowledgments:}} We thank Zbis{\l}aw
Tabor for making his code available to us. The first two authors
acknowledge hospitality of the AEI in Golm during part of this work.
The research of PB and TC was supported in part by the Polish
Research Committee grant 1PO3B01229.

\end{document}